\documentclass[%
 aip,
 jmp,%
 amsmath,amssymb,
 reprint,%
]{revtex4-1}
\usepackage{mathrsfs}
\usepackage{graphicx}
\usepackage{dcolumn}
\usepackage{bm}

\begin{document}


\title{The Influence of s-wave interactions on focussing of atoms}

\author{A. M. Kordbacheh}

%
\affiliation{ 
	Department of Quantum Science, Research School of Physics, The Australian National University, Canberra, ACT 2601 Australia
}
\author{A. M. Martin}%
\affiliation{ 
	School of Physics, University of Melbourne, Melbourne, 3010, Australia
}%

\date{\today}

\begin{abstract}
The focusing of a rubidium Bose-Einstein condensate via an optical lattice potential is numerically investigated. The results are compared with a classical trajectory model which under-estimates the full width half maximum of the focused beam. Via the inclusion of the effects of interactions, in the Bose-Einstein condensate, into the classical trajectory model we show that it is possible to obtain reliable estimates for the full width half maximum of the focused beam when compared to numerical integration of the Gross-Pitaevskii equation. Finally, we investigate the optimal regimes for focusing and find that for a strongly interacting Bose-Einstein condensate focusing of order $20$ nm may be possible.
\end{abstract}

\maketitle

\section{Introduction}
\label{sec:1}

Atom lithography is a technique where the gradient forces applied by laser fields on a beam of atoms are used to direct the atoms into nanostructures deposited on a plane surface\cite{p1,p2}. It has been the topic of considerable study in the field of atom optics over the last two decades as it provides a scheme of writing nanometer structures directly onto a substrate in parallel process \cite{p3}. The main application in the development of nano-lithography could be the race for an increased density of transistors in the computer chips \cite{p54}. The possibility of using light to deposit feature sizes of a few nanometers was first suggested in 1987 by Balykvin and Letokhov \cite{p44,p45}. Later in 1991, McClelland and Scheinfein \cite{p46} proposed a particle optics approach in which the atom lens created by the laser light can focus an atomic beam down to a surface. The principle was experimentally demonstrated by Timp \cite{p1} in 1992 using Na deposited on Si via a standing light wave, and was similarly followed by McClelland et al. \cite{p2} to focus Cr in the presence of a 1D lattice. Atom lithography continued to develop to two- and three-dimensional nanostructures in a single process utilizing the selectivity of atom-light interaction (using 2D or 3D lattices) \cite{p49}. There have also been demonstrations depositing Yb \cite{p50} and Fe \cite{p51,p52}. Almost all the experiments accomplished so far, have used an oven source of atoms in which the beam is collimated with an aperture followed by a transverse laser cooling process\cite{p2} before traveling through a focusing potential. The traditional study of focusing in optical lattices uses the classical trajectories model of atomic motion when travelling through the light \cite{p46}. The principle is based on atom-light interactions, resulting from the dipole force \cite{p7} which causes neutral atoms to become manipulated with a near resonant laser light \cite{p7,p8,p9}. Ideally, this model predicts almost perfect focusing in the absence of interactions.

Nevertheless, there are some advantages in using a Bose-Einstein condensate (BEC) of neutral atoms as the source of atom deposition. Since the de-Broglie wavelength of a gas of atoms is of the order of the mean field distance between particles, an ultra cold source such as a BEC would bring atoms to wavelengths of $\sim$ nm or pm for nano-Kelvin temperatures \cite{p4} resulting in an excellent collimation of the beam of atoms as well as a high flux density \cite{p5}. Using a BEC source can also reduce effects such as chromatic aberration and angular divergence \cite{p2}, with the longitudinal and transverse velocity distributions typically being much lower when incident on the surface compared to those resulted from thermal sources.

However, for BEC's, the effect of interactions must be accounted for by introducing the s-wave interaction between atoms. The investigation of the matter wave focusing dynamics of a trapped BEC for both an interacting and non-interacting case was conducted theoretically by D. Muuray and P. Ohberg in 2004 \cite{p53}. In their work, they derived the time to focus as a function of the focusing strength for a 3D Bose-condensed cloud. The significance of this research is to take atomic interactions into account when focusing a free propagating BEC and to scale its effect on the broadening of the focal spot sizes and peak densities achievable in realistic nano-lithography experiments. 

\begin{figure}[t!]
	\centering
	\includegraphics[width=9cm, height=8cm,angle=0] {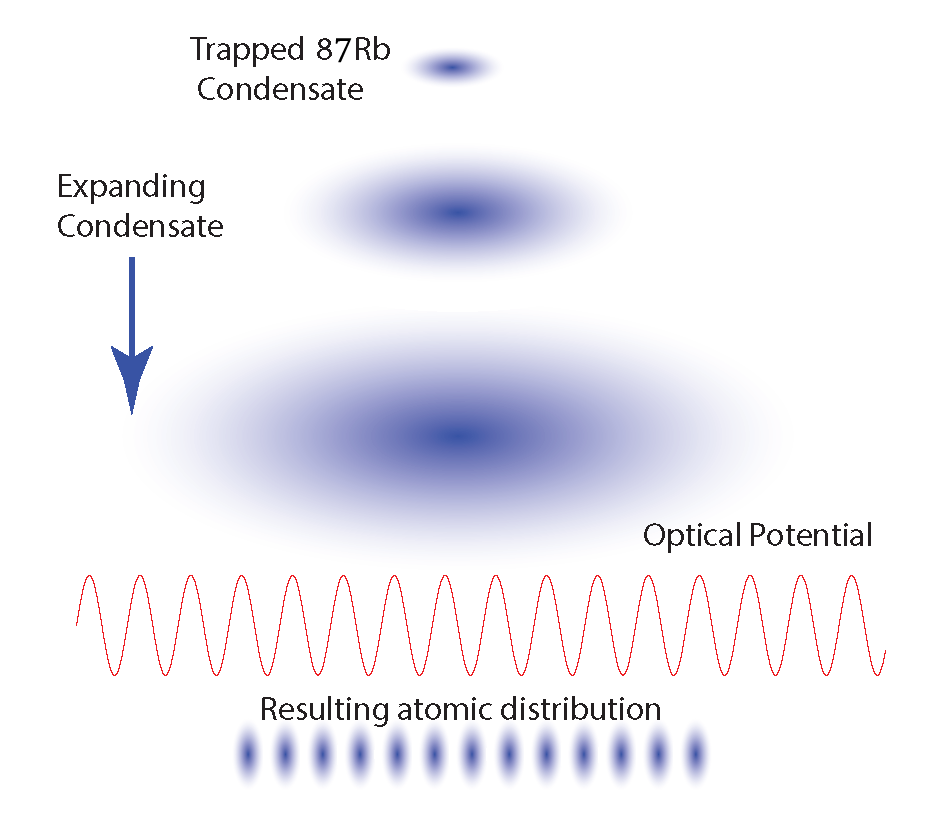}
	\caption{Schematic representation of atom deposition using a $^{87}$Rb BEC focused by an optical lattice. The condensate is initially confined by a harmonic trap. Once the BEC is released from the trap, it starts propagating whilst expanding along the falling (longitudinal) direction. An optical potential with a sinusoidal configuration along the transverse axis focuses the BEC resulting in the periodic atomic distribution.}
	\label{f1}
\end{figure}

The principle of atom lithography using a BEC is schematically depicted in Fig~\ref{f1}. The cloud of $^{87}$Rb is initially confined by a harmonic trap. Turning off the trap, the released condensate expands whilst propagating along the vertical axis. It then encounters the optical lattice being focused by its nodes along the horizontal axis resulting in a large periodic array of nano-structures that are separated by $\lambda /2$ where $\lambda$ is the wavelength of lattice.

In this paper, the focal properties of an optical lattice are derived by the classical equations. Using this model, we estimate the Full Width at Half Maximum (FWHM) as well as peak density of $^{87}$Rb focused structures, which can be further used to determine the possible contributions of structure broadening such as the spherical and diffraction aberrations. Following this, the Gross-Pitaevskii Equation (GPE) is introduced and its application for the purpose of BEC lithography is studied. Conducting various numerical simulations via the GPE for different $^{87}$Rb BEC and focusing potential parameters, we estimate resultant focused structure resolutions and peak densities. Not only is the key role of the s-wave interaction within the BEC investigated through the GPE, but also by taking the transverse and longitudinal velocity distributions of the condensate, the classical trajectories model is developed to consider the influence of atomic interactions on focused profiles.

\section{The Classical Trajectories Approach}
\label{sec:2}

When atoms are exposed to a potential field, an atomic dipole moment is induced by the electric field. The induced dipole moment interacts with the gradient of the field resulting in a dipole force gradient \cite{0} applied towards the nodes or anti-nodes of the periodic field. The resultant dipole potential on stationary atoms is well established \cite{0}. However, when atoms carry a momentum, the created optical dipole potential behaves as a focusing thick lens on neutral moving atoms \cite{17}:
\begin{equation}
\begin{split}
U(x,z)=\frac{\hbar\Delta}{2}\ln(1+p(x,z));\\ p(x,z)=\frac{\gamma^2}{\gamma^2+4\Delta^2} \frac{I(x,z)}{I_s},\quad
\label{1}
\end{split}
\end{equation}
where $\Delta$ denotes the detuning of the laser frequency from the atomic resonance, $\hbar$ is the Plank constant, $\gamma$ is the natural linewidth of the atomic transition (spontaneous decay rate from the excited level) and $I_s$ is the saturation intensity related to the atomic D$_2$ line transition, $5~^2S_{1/2}\longrightarrow 5~^2P_{3/2}$, for $^{87}$Rb. While a red detuned laser light from resonance, $\Delta<0$, in $^{87}$Rb D$_2$ line, directs the falling atoms towards the nodes of the standing potential, a blue detuned light, $\Delta>0$ brings the atoms to the anti-nodes of the focusing potential ($\Delta=\omega_L-\omega_0$ where $\omega_L$ and $\omega_0$ are respectively the frequency of incident laser light and resonance between $5~^2S_{1/2}$ and $5~^2P_{3/2}$). We consider a geometry for the optical potential such that the laser intensity profile, $I(x, z)$, shapes a mask with a periodic scheme of multiple focusing lenses along the direction of focusing atoms. Hence, a Gaussian standing potential is considered which contains of a sinusoidal behavior of the intensity along the $x$-axis (the focusing direction) and a Gaussian envelope function along the $z$-axis, the direction of falling atoms. This optical lattice potential is represented by
\begin{equation}
I(x,z)=I_0 \exp(-2z^2/\sigma_z^2)\sin^2(kx),
\label{2}
\end{equation}
where $I_0$ is the maximum intensity of the spatially varying Gaussian profile, $\sigma_z$ is the radius of the beam at $1/e^2$ value of the maximum intensity and $k=2\pi/\lambda$ is the wavenumber. Since almost no force is applied to atoms along the $y$ direction compared to the other two directions, neglecting this causes the focusing potential, Eq.~(\ref{1}), to become independent of $y$.

In order to scale the parameters involved in an optimal focusing potential, we exploit the classical trajectories approach \cite{p2}. Neglecting the $y$ axis due to the symmetry of problem, the classical equations of motion for atomic trajectories are given by
\begin{equation}
\frac{d^2x}{dt^2}+\frac{1}{m}\frac{\partial U(x,z)}{\partial x}=0;
\label{3}
\end{equation}
\begin{equation}
\frac{d^2z}{dt^2}+\frac{1}{m}\frac{\partial U(x,z)}{\partial z}=0.
\label{100}
\end{equation}
Using conservation of energy, one can combine Eqs.~(\ref{3}) and (\ref{100}) solving for $x$ as function of $z$
\begin{equation}
\begin{split}
\frac{d}{dz}\bigg[\bigg(1-\frac{U(x,z)}{E_0}\bigg)^{1/2}\big(1+x^{\prime2}\big)^{-1/2}x^\prime\bigg]~~~~~~~~~~~~~~~~\\
+\frac{1}{2E_0}\bigg(1-\frac{U(x,z)}{E_0}\bigg)^{-1/2}\big(1+x^{\prime2}\big)^{1/2}\frac{\partial U(x,z)}{\partial x}=0,
\label{206}
\end{split}
\end{equation}
where $E_0$ represents the total energy of each atom and $x^\prime=\frac{dx}{dz}$.

To obtain the focal properties of the lattice, we consider the paraxial approximation \cite{p20} in which the sinusoidal term of the potential consisting of a multi-node wave along the $x$ axis is converted to a single node harmonic part. In addition, the approximation neglects aberrations, i.e.  the trajectories are assumed to be perfectly parallel to the $z$ axis when falling towards the lattice. These can be implemented by $U(x,z)\ll E_0$, $\frac{dx}{dz} \ll 1$ and $kx\ll 1$. Applying these, Eq.~(\ref{206}) converts to  
\begin{equation}
\frac{d^2x}{dz^2}+\frac{1}{2E_0}\frac{\hbar\Delta}{2}\bigg[\frac{1}{1+p(x,z)}\frac{\partial p(x,z)}{\partial x}\bigg]=0.
\label{210}
\end{equation}
Now considering $\sin(kx)\sim kx$, $\sin^2(kx)\sim 0$ and $\cos(kx)\sim 1$, one obtains the following second order differential equation
\begin{equation}
\frac{d^2x}{dz^2}+q^2\exp(-2z^2/\sigma_z^2)x=0,
\label{4}
\end{equation}
\begin{equation}
q^2=\frac{\hbar\Delta}{2E_0}\frac{I_0}{I_s}\frac{\gamma^2}{\gamma^2+\Delta^2}k^2.
\label{213}
\end{equation}
According to the relation between the maximum intensity and the corresponding value of power in a standing wave Gaussian beam \cite{p20}, $I_0=8P_0/\pi\sigma_z^2$, the required power value of the harmonic potential to focus atoms at any desired spots along the focal axis ($z$-axis) is achieved as function of the potential factors and atoms' initial kinetic energy,
\begin{equation}
P_0=\xi\frac{\pi}{4}\frac{E_0}{\hbar\Delta}\frac{\gamma^2+4\Delta^2}{\gamma^2}\frac{I_s}{k^2},
\label{5}
\end{equation}
where $\xi=q^2\sigma_z^2$ is a dimensionless parameter.

\begin{figure}[tb]
	\centering
	\hskip 2ex
	\includegraphics[width=8.5cm, height=8.5cm,angle=0]{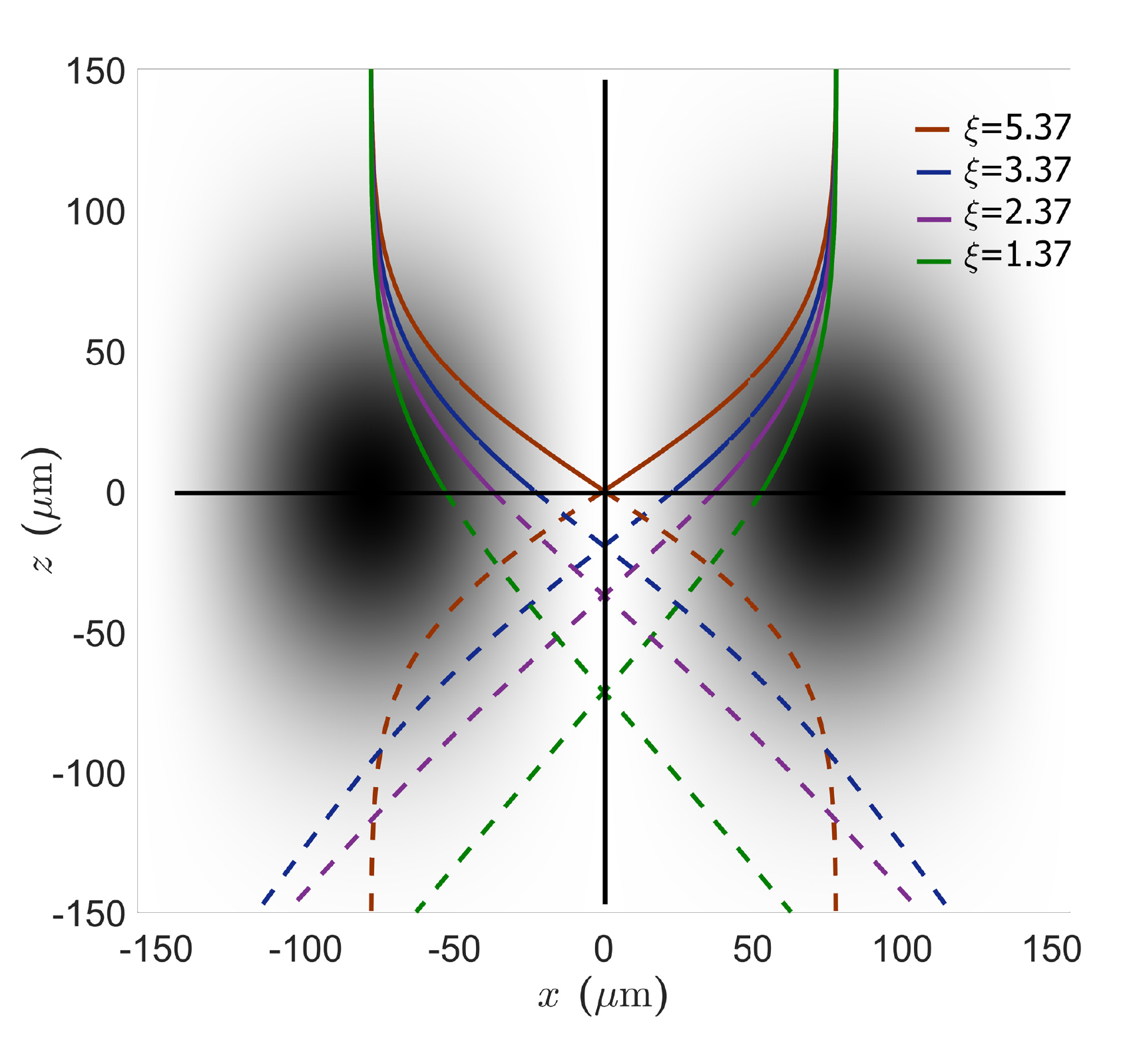}
	\caption{Cross-section view (top view) for the $^{87}$Rb classical atomic trajectories for different values of $\xi$ and consequently different harmonic potential powers, $P_0$, when applying the paraxial approximation. The solid and dashed curves respectively represent the area above ($z>0$) and below ($z<0$) the center of potential ($z=0$). The two gray-shaded oval areas illustrate the harmonic potential intensity profile,  $I(x, z)$, which are darker for higher intensity regions. The maximum intensity value for $\xi=5.37$, $3.37$, $2.37$, and $1.37$ is respectively $I_0=1.095\times 10^4$, $6.874\times 10^3$,$4.834\times 10^3$, and $2.794\times 10^3$ W/m$^2$. A choice of $\lambda=400\lambda_{\text{D}_2}=312~\mu$m for the potential wavelength necessitates the two adjacent peaks (located at $x=\pm\lambda /4=\pm78~\mu$m, $z=0$) to be apart by $\lambda/2=156~\mu$m along the $x$ axis.
		The trajectories corresponding to $\xi=5.37$ (brown curves) are focused at $x_f=z_f=0$ while setting $\xi=3.37$, $2.37$, and $1.37$ shifts the focus points to $z<0$. 
		Parameters used in the simulation are:  $\sigma_z=100\mu$m, $\Delta=200$ GHz, $\gamma=37$ MHz, $I_s=16.5$ W/m$^2$, $m=1.44\times 10^{-25}$ Kg and $v_i=v_z=1$ cm/s.}
	\label{f2}
\end{figure}

\begin{figure}[tb]
	\hskip -8ex
	\includegraphics[width=9.5cm, height=10cm,angle=0]{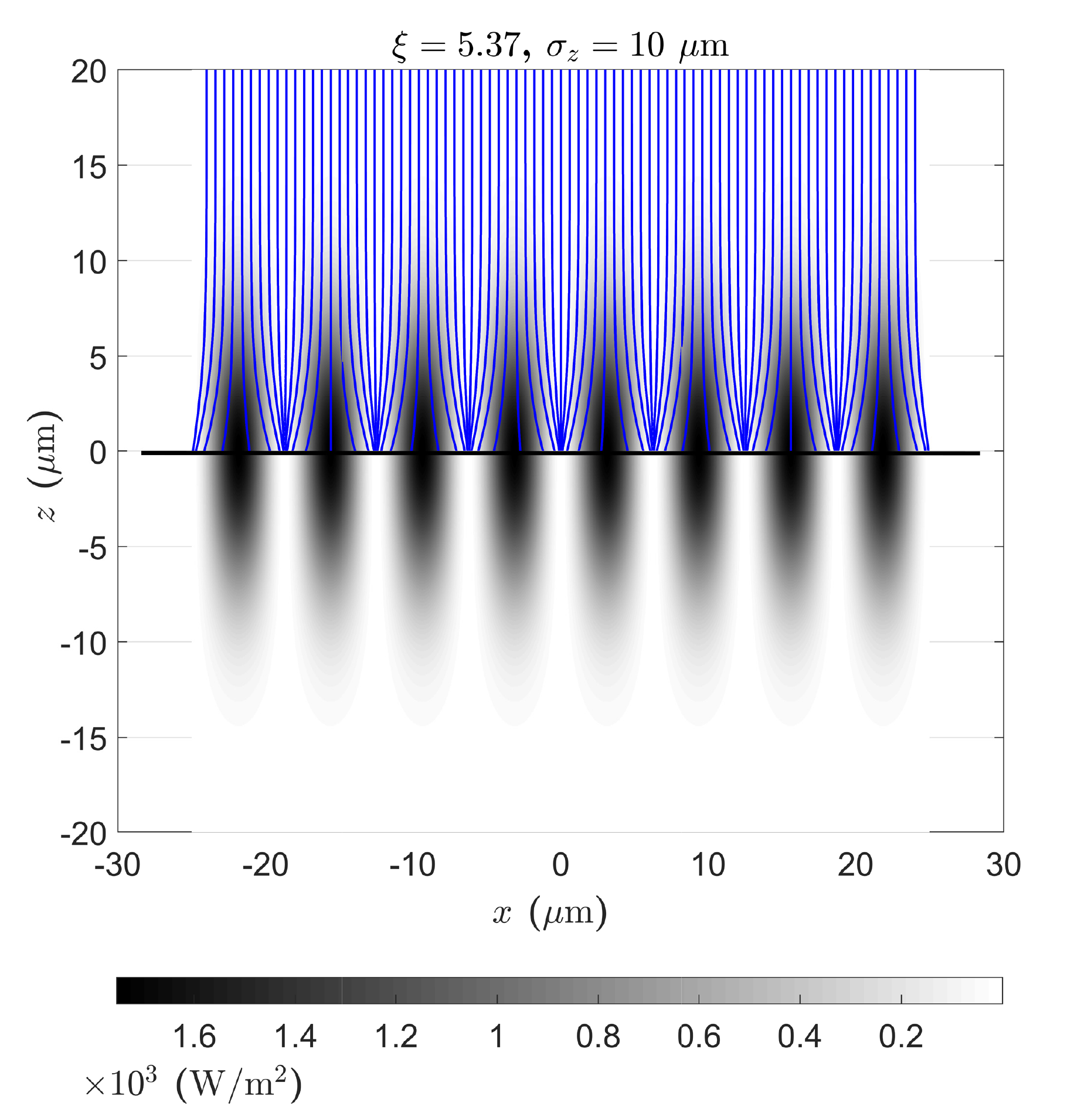}
	\caption{Top view of the classical trajectories for the $^{87}$Rb atoms (indicated by the solid blue curves) falling at $v_z=1$ cm/s between $x=-24~\mu$m and $x=24~\mu$m being deposited at $z_f=0$, and the lattice (depicted by the gray-shaded oval areas) of a size of $\sigma_z=10~\mu$m distributed between $x=-8\lambda/4=-24.961~\mu$m and $x=8\lambda/4=24.961~\mu$m. The color map at the bottom of the figure represents the intensity of the focusing lattice, and the horizontal solid black line shows the focal plane. The lattice power and maximum intensity values are $P_0=0.068~\mu$W and $I_0=1.752\times 10^3$ W/m$^2$ given that $\xi=5.37$ and $\Delta=200$ GHz.
		Parameters involved in the simulation are: $\gamma=37$ MHz, $I_s=16.5$ W/m$^2$, $m=1.44\times 10^{-25}$ Kg.}
	\label{f3}
\end{figure}

\begin{figure}[ht]
	\centering
	\includegraphics[width=9cm, height=8.5cm,angle=0]{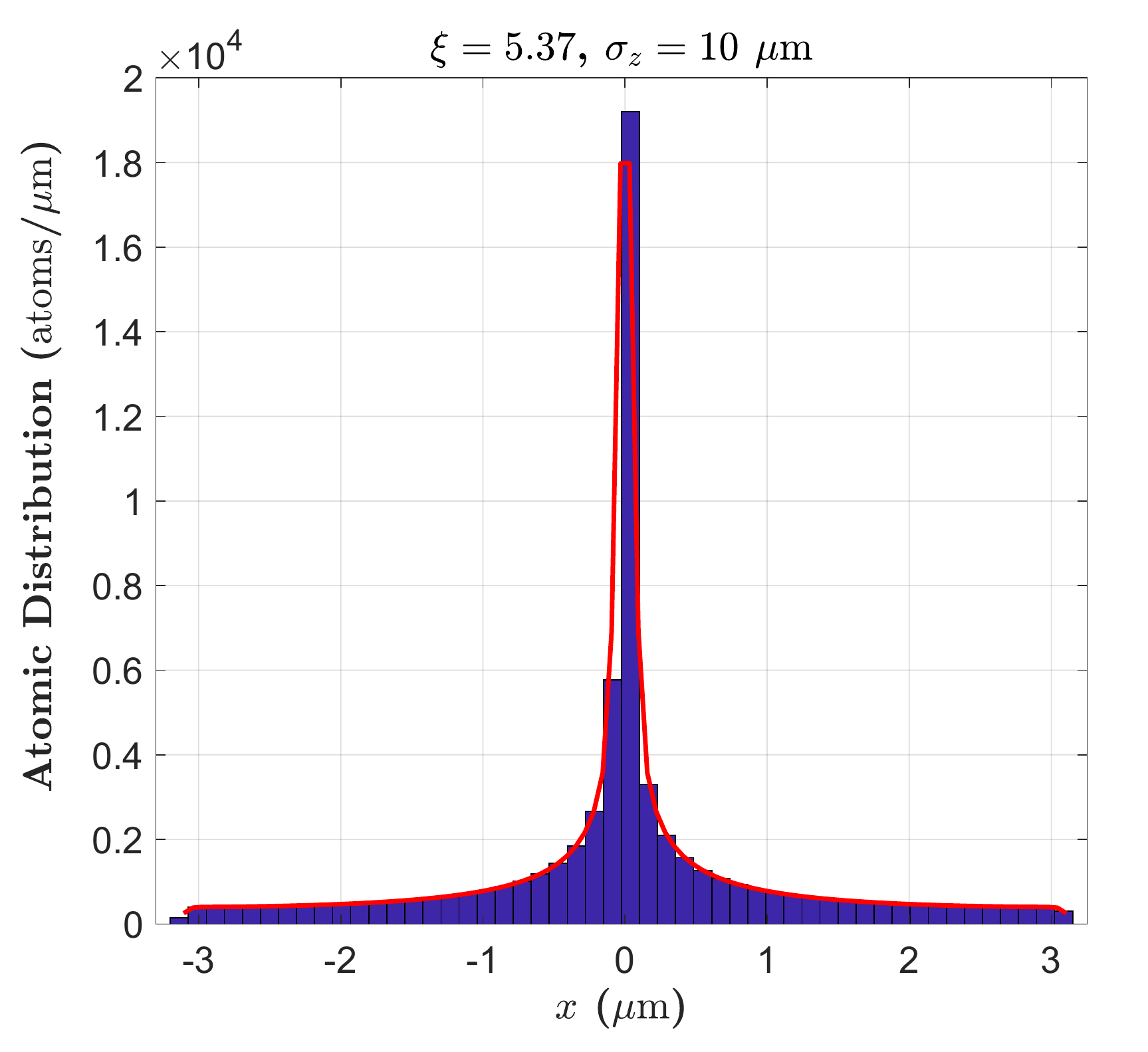}
	\caption{Histogram distribution indicated by the violet area including $n=50$ bins show the 62400 atoms focused on the focal plane, $z_f=0$, between $x=-\lambda /4=-3.12~\mu$m and $x=\lambda /4=3.12~\mu$m. The estimated FWHM is $(\Delta x)_{\text{sph}}=0.136~\mu$m using a Kernel fit to the distribution illustrated by the red solid curve. Parameters used in the simulation are:  $\sigma_z=10~\mu$m, $\Delta=200$ GHz, $\gamma=37$ MHz, $I_s=16.5$ W/m$^2$, $m=1.44\times 10^{-25}$ Kg, $v_z=1$ cm/s and $\lambda=12.48~\mu$m.}
	\label{f4}
\end{figure}

We now proceed with a specific example for focusing ultra-cold $^{87}$Rb atoms using the paraxial approximation displayed in Fig.~\ref{2}. To meet the requirements of approximation, the potential wavelength is chosen to be 400 times greater than the actual $^{87}$Rb D$_2$ line wavelength, $\lambda=400\lambda_{\text{D}_2}=312~\mu$m where $\lambda_{\text{D}_2}=780.027$ nm. This forms a potential with a harmonic distribution along the $x$ axis. While the potential radius is adjusted to $\sigma_z=100~\mu$m, $\xi=5.37$, $3.37$, $2.37$, and $1.37$ corresponding to the potential power of $P_0=43.018$ , $26.996$, $18.986$, and $10.975~\mu$W respectively provided that the initial velocity of atoms and the detuning from resonance are selected as $v_i=v_z=1$ cm/s and $\Delta=200$ GHz. 
For each value of $\xi$ or $P_0$, two $^{87}$Rb atomic trajectories falling symmetrically from  $x_i=\pm\lambda /4=\pm78~\mu$m landing at $x_f=0$ and a particular $z_f$ are plotted. As a case in point, for $\xi=5.37$, solving Eq.(\ref{4}) yields the focal point optimally placed at $z_f=0$ and $x_f=0$ (the center of potential). However, reducing power values results in the focal point being located at $z<0$ below the potential center. 

We note that this is an ideal scenario in which for a particular value of $P_0$, all atoms are focused exactly at one spot and no aberration is considered. However, in practice there always exists a spherical aberration in optics while atoms are entering a thick lens \cite{19}. This arises from the fact that atoms traveling farther to the focal axis will experience a weaker force than those that are closer to this axis. This effect can be considered through the exact numerical solution of  Eq.(\ref{206}). As an illustration, the atomic trajectories for ultra-cold $^{87}$Rb starting at $z_i=20~\mu$m whilst moving at a longitudinal velocity of $v_z=1$ cm/s through the potential of a radius size of $\sigma_z=10~\mu$m focused at $z_f=0$ ($\xi=5.37$) have been numerically simulated and are depicted in Fig~\ref{f3}. Here, we choose a lattice whose wavelength is 16 times larger than the actual wavelength of $^{87}$Rb D$_2$ line, $\lambda=16\lambda_{\text{D}_2}=12.48~\mu$m. This is practical in realistic experiments through the use of a Spatial Light Modulator (SLM) \cite{30, 31}. Aligning the laser detuning to $\Delta=200$ GHz, the required optimal value of lattice power and maximum peak intensity to focus the atoms to the center of potential ($z=0$) are calculated as $P_0=0.068~\mu$W and $I_0=1.752\times 10^3$ W/m$^2$. Unlike the paraxial solution, for each lattice node here, the atoms do not fully land at the same focal point, and this process is identical for all lattice nodes implying the spherical aberration. Hence, one could evaluate the linewidth of the created structure and estimate the broadening contribution arising from this effect inside every lattice slit, $D=\lambda/2=6.24~\mu$m. For instance, we have selected the lattice central node, from $x=-\lambda/4=-3.12~\mu$m to $x=\lambda/4=3.12~\mu$m, and plotted a histogram distribution for 62400 trajectories arriving at the focal plane ($z=0$) in Fig~\ref{f4}. Applying a Kernel fit to the focused profile, the value of FWHM is acquired as $(\Delta x)_{\text{sph}}=0.136~\mu$m.

However, the spherical aberration is not the only reason for profile broadening. In a lens like lattice, there are finite adjacent apertures positioned in a distance of $\lambda/2$ apart along the entire lens. Since atoms exhibit wave-like behavior, they interfere together with their de-Broglie wavelengths, $\lambda_{\text{dB}}$, while crossing through the apertures. This causes a limit on the linewidth of structures known as the diffraction effect. The angular resolution produced by the diffraction is estimated by Rayleigh Criterion \cite{2_13, 2_14}
\begin{equation}
\theta=\beta\frac{\lambda_{dB}}{D},
\label{223}
\end{equation}
where the factor $\beta=1.22$ accounts for a circular aperture \cite{2_13} while $\beta=0.88$ is dedicated to a rectangular (or cylindrical) aperture \cite{2_15, 2_16, p20}.  The de-Broglie wavelength of atoms is defined as $\lambda_{\text{dB}}={h}/{mv_z}$ where $v_z$ is the most probable longitudinal velocity. The angular resolution in Eq.~(\ref{223}) can be converted to the diffraction-limited FWHM by
\begin{equation}
(\Delta x)_\text{diff}=f\tan(\theta)\approx f\theta,
\label{225}
\end{equation}
where $\theta$ is considered to be a very small angle. Since each aperture has a rectangular structure to the incident atomic beam, $(\Delta x)_\text{diff}$ would be
\begin{equation}
(\Delta x)_\text{diff}=0.88 \frac{f\lambda_{\text{dB}}}{D}=1.76\frac{fh}{mv_z\lambda}.
\label{226}
\end{equation}
For the previous example, for the atoms with a mass of $m=1.44\times 10^{25}$ Kg falling at $v_z=1$ cm/s through a lattice of wavelength of $\lambda=312~\mu$m, the diffraction-limited FWHM is estimated as $(\Delta x)_\text{diff}=0.3589~\mu$m. In this calculation, the focal length (the difference between the coordinates of the focal point and the principal plane location when focusing at $z=0$), $f=5.531~\mu$m, $\sigma_z=10~\mu$m and $\lambda=12.48~\mu$m. According to Eq.~(\ref{226}), this broadening is proportionally related to the focal length and is inversely associated with the velocity of atoms as well as the size of apertures. Hence, to minimize $(\Delta x)_\text{diff}$, one needs to either use a relatively shorter focal length, a higher longitudinal velocity or a larger potential wavelength. 

Eventually, taking into account the total structure broadening predicted by the classical trajectories model in absence of atomic interactions,
\begin{equation}
(\Delta x)_\text{classical}=(\Delta x)_\text{sph}+(\Delta x)_\text{diff},
\label{227}
\end{equation}
and considering the results for $(\Delta x)_\text{sph}$ of the stated example, we can estimate the total value of FWHM via the classical trajectories model for the cases, $z_f=0$ (with $\sigma_z=10~\mu$m), which is evaluated as $(\Delta x)_{\text{Classical}}=0.495~\mu$m.

\begin{figure*}
	\hskip -5ex
	\includegraphics[width=14.5cm, height=15.5cm,angle=0]{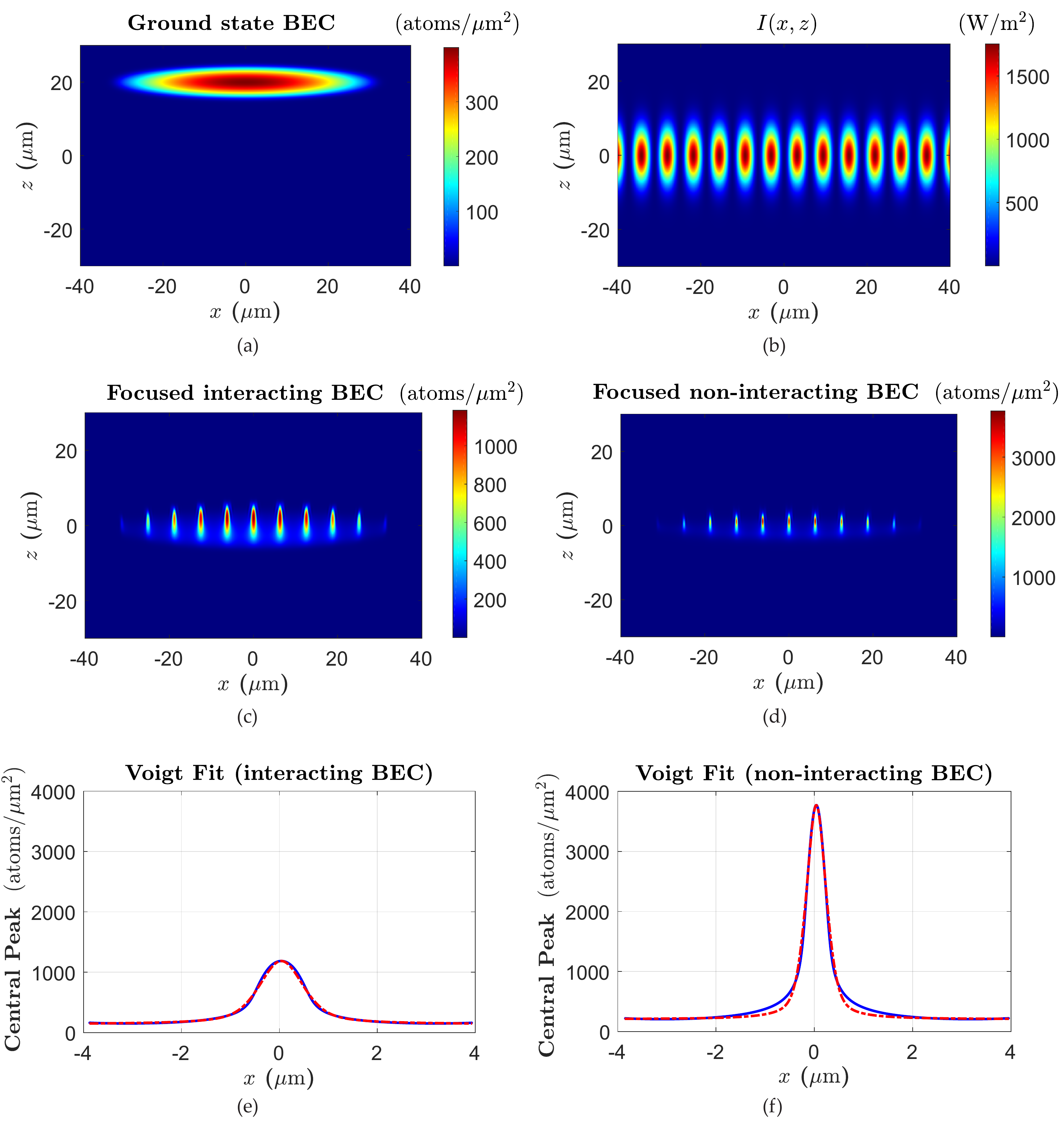}
	\caption{(a): Cross section view ($x-z$ plane) of $^{87}$Rb BEC ground state. The center-of-mass of the condensate is located at $z_0=20~\mu$m. (b): Intensity profile of the lattice potential in W/m$^2$ indicated by the color map. (c), (d): Focused structures, respectively, for an interacting and non-interacting $^{87}$Rb BEC from the prospective of $x-z$ plane. (e), (f): The transverse profile of the central peak along the $x$ axis for the interacting and non-interacting BEC (solid blue curves) whose linewidth is estimated as $(\Delta x)_{\text{GPE}}^{\text{int}}=1.074~\mu$m  and $(\Delta x)_{\text{GPE}}^{\text{non}}=0.477~\mu$m using a Voigt fit (dashed red curves). The BEC and intensity profiles in (a)-(d) have been integrated over the $y$ axis. For this example, we assume that the BEC moves with a constant velocity under no gravity, $g=0$. Parameters involved in the simulations are: $\xi=5.37$, $N=10^5$, $\omega_x=2\pi\times10$ Hz, $\omega_y=\omega_z=2\pi\times70$ Hz, $\lambda=12.48~\mu$m, $I_s=16.5$ W/m$^2$, $\Delta=200$ GHz, $\sigma_z=10~\mu$m, $v_z=1$ cm/s, $m=1.44\times 10^{-25}$ Kg, and $a_s=100a_0$ and $0$ for the interacting and non interaction cases respectively.}
	\label{f303}
\end{figure*}

\section{The Gross-Pitaevskii Equation Methodology}
\label{sec:3}

 Using a BEC \cite{3_1} as a source of ultra-cold atoms brings several advantages to atom deposition as it can significantly reduce the linewidth of longitudinal and transverse velocity distributions providing excellent coherence and collimation for the atomic beam as well as offering relatively small de Broglie wavelengths, high peak densities and quality spatial modes \cite{p4, p5, p12}. In this section, we take into account the atomic interaction when focusing a free propagating $^{87}$Rb BEC. The impact of interactions on the broadening of the nano-focal spot sizes and peak densities are estimated, which is accomplished through the use of the GPE \cite{3_8, 3_9}.

The three-dimensional time dependent GPE modeling the dynamics of a BEC is represented by \cite{3_8, 3_11, 3_12, 3_13}
\begin{equation}
i\hbar\frac{\partial\psi(\mathbf{r},t)}{\partial t}=\Big(-\frac{\hbar^2}{2m}\nabla^2+V_{\text{ext}}(\mathbf{r},t)+V_{\text{mean}}(\mathbf{r},t)\Big)\psi(\mathbf{r},t),
\label{301}
\end{equation}
where $\psi(\mathbf{r},t)$ indicates the BEC wavefunction at different times of propagation, $V_{\text{ext}}$ is the time-dependent external potential applied on the BEC, $m$ is the atomic mass and $\hbar$ is the Planck constant. The interactions between atoms within the cloud are considered using a non-linear mean field potential, $V_{\text{mean}}$, which estimates the averaged exerted potential on any particular atom by all other atoms given by \cite{p10, 3_15, p12, 3_17}
\begin{equation}
V_{\text{mean}}(\mathbf{r},t)=u|\psi(\mathbf{r},t)|^2,
\label{302}
\end{equation}
where
\begin{equation}
u=\frac{4\pi\hbar^2a_s}{m},
\label{303}
\end{equation}
quantifies the atomic interactions, $|\psi(\mathbf{r},t)|^{2}$ describes the atomic density, and $a_s$ is the s-wave scattering length. The value $a_s$ can be practically tuned from $a_s>0$ (repulsive interactions) to $a_s<0$ (attractive interactions) utilizing Feshbach resonance \cite{3_19}.

\subsection{The BEC Ground State}
\label{sec:4}

In order to generate the BEC ground state wavefunction at $t=0$, we assume that the condensate is initially confined via a harmonic trapping potential defined by
\begin{equation}
V_{\text{ext}}(\mathbf{r}, 0)=V_{\text{trap}}(\mathbf{r})=\frac{1}{2}m(\omega_x^2x^2+\omega_y^2y^2+\omega_z^2z^2),
\label{304}
\end{equation}
where $\omega_x$, $\omega_y$, $\omega_z$ represent the harmonic trap frequencies along the $x$, $y$, $z$ axes respectively. For our purpose, we consider a cylindrical (cigar-shaped) condensate with two radial axes associated with the two tight trap frequencies, $\omega_y$ and $\omega_z$, and one axial axis corresponding to the weak trap frequency, $\omega_x$, where $\omega_y$, $\omega_z>\omega_x$. This configuration allows the BEC to be elongated along the $x$ axis compared to the $y$ and $z$ axes. 

The Thomas-Fermi solution \cite{3_20, 3_18} to Eq.~(\ref{301}) can be used as an initial function to Eq.~(\ref{301}) to acquire the exact solution for the ground state wavefunction of system, $\psi_g(\mathbf{r}, t=0)$, which itself is exploited as an initial condition to Eq.~(\ref{301}) to achieve the propagation state, $\psi(\mathbf{r}, t>0)$. The process of calculating $\psi_g(\mathbf{r}, t=0)$ and $\psi(\mathbf{r}, t>0)$ is numerically conducted using an imaginary and real time step respectively via the Embedded Runge-Kutta scheme along with adaptive Fourier split-step size \cite{3_21}. We consider  $10^{5}$ atoms trapped by $\omega_x=2\pi\times10$ Hz and $\omega_{y, z}=2\pi\times70$ Hz producing a cylindrical BEC elongated along the $x$ axis. Once the harmonic trap, $V_{\text{trap}}$, is turned off ($\omega_i=0$, $i=x, y, z$), the condensate starts expanding due to the s-wave interaction between atoms. 

\begin{table*}[ht!]
	\centering
	\footnotesize
	\begin{tabular}{|c |c|c|cc|cc|} 
		\hline
		Example & $P_0$ ($\mu$W) & $I_0$ (W/m$^2$) & \parbox[t]{1.5cm}{\raggedright $(\Delta x)_{\text{GPE}}^{\text{int}}$ ($\mu$m)} & \parbox[t]{1.2cm}{\raggedright $(\Delta x)_{\text{GPE}}^{\text{non}}$ ($\mu$m)} & \parbox[t]{2cm}{\raggedright (Peak)$_{\text{GPE}}^{\text{int}}$ (atoms/$\mu$m$^2$)} & \parbox[t]{2cm}{\raggedright (Peak)$_{\text{GPE}}^{\text{non}}$ (atoms/$\mu$m$^2$)}\\
		
		\hline
		\parbox[t]{2cm}{\raggedright $\sigma_z=10~\mu$m $v_z=1$ cm/s} & $0.0688$ & $1.752\times 10^3$ & $1.074$ & $0.477$ & $1185$ & $3771$ \\
		\hline
		\parbox[t]{2cm}{\raggedright $\sigma_z=20~\mu$m $v_z=1$ cm/s}  & $0.0688$ & $4.381\times 10^2$ & $2.106$ & $0.985$ & $543$ & $1598$ \\
		\hline
		\parbox[t]{2cm}{\raggedright $\sigma_z=10~\mu$m $v_z=2$ cm/s} & $0.275$ & $7.01\times 10^3$ & $0.447$ & $0.228$ & $3181$ & $6028$ \\
		\hline
	\end{tabular}
	\caption{FWHM and peak density results for interacting ($a_s=100a_0$) and non-interacting ($a_s=0$) focused BEC's collected at $z=0$ for three different simulations \{$\sigma_z=10~\mu$m, $v_z=1$ cm/s\}; \{$\sigma_z=20~\mu$m, $v_z=1$ cm/s\} and \{$\sigma_z=10~\mu$m, $v_z=2$ cm/s\}.}
	\label{t301}
\end{table*}

\subsection{The Time Dependent Focusing Potential}
\label{sec:5}

For simplicity of numerical calculation for the evolving BEC through a focusing potential, we assume that the BEC is located in a stationary frame while the optical lattice is situated in a moving frame approaching the BEC along the $z$ axis. Hence, $V_{\text{lattice}}$ would be dependent on time by $z(t)=\frac{1}{2}gt^2+v_0t$, which is the varying distance as a function of time following the free falling method where $g$ and $v_0$ are the gravity and initial velocity kick respectively. Thus, once the confining potential is switched off at $t>0$, the optical lattice potential [see Eq.~(\ref{1})] is switched on so that
\begin{equation}
\begin{split}
V_{\text{ext}}(\mathbf{r}, t)=V_{\text{lattice}}(x, t)=~~~~~~~~~~~~~~~~~~~~~~~~~~~~~~~~~~~~~~~~~~\\
\frac{\hbar\Delta}{2}\ln\Bigg[1+\frac{I_0}{I_s}\frac{\gamma^2}{\gamma^2+4\Delta^2}\exp\Bigg(\frac{-2\big(z_0-z(t)\big)^2}{\sigma_z^2}\Bigg) \sin^2(kx)\Bigg],
\label{309}
\end{split}
\end{equation}
where $z_0$ denotes the initial distance between the center of lattice and the center-of-mass of condensate along the $z$ axis. As a result, combining Eqs.~(\ref{301}), (\ref{302}), (\ref{303}), and (\ref{309}), the dynamics of a focused BEC at $t>0$ would be given by the following equation
\begin{equation}
	\begin{split}
i\hbar\frac{\partial\psi(\mathbf{r},t)}{\partial t}=\Big(-\frac{\hbar^2}{2m}\nabla^2+V_{\text{lattice}}(x, t)\\
+\frac{4\pi\hbar^2a_s}{m}|\psi(\mathbf{r},t)|^2\Big)\psi(\mathbf{r},t).
\label{310}
\end{split}
\end{equation}

To make a direct comparison between the output of the GPE and classical trajectories models, we exploit the same example in section~\ref{sec:2} in which the $^{87}$Rb BEC located at $z_0=20~\mu$m is released from the trap by $v_i=v_z=1$ cm/s [see Fig~\ref{f303}(a)] optimally focusing to the center ($\xi=5.37$, $z=z_f=0$) of the lattice potential with $\sigma_z=10~\mu$m and  $\lambda=16\lambda_{\text{D}_2}=12.48~\mu$m [see Fig~\ref{f303}(b)]. For this case, provided that the laser detuning is set to $\Delta=200$ GHz, the optimal lattice power and maximum peak intensity values to focus the BEC at $z=0$ are required as $P_0=0.068~\mu$W and $I_0=1.752\times 10^3$ W/m$^2$ respectively. Applying the stated factors along with the relevant data for the mass, saturation intensity and spontaneous emission rate associated with $^{87}$Rb D$_{2}$ line in Eq.~(\ref{310}), the BEC focusing dynamics is numerically achieved, and the full process as well as results are shown in Figs~\ref{f303}(a-f).

Here we have considered two different cases to examine the focal spot sizes and peak densities. Firstly, the BEC is strongly interacting and the s-wave scattering length is chosen as $a_s=100a_0$ leading to a repulsive BEC [see Fig~\ref{f303}(c)]. Secondly, it is assumed that there exists no atom-atom interactions within the cloud, $a_s=0$ [see Fig~\ref{f303}(d)] in the focusing process enabling one to compare directly the outcomes with those of the spherical aberration from the classical trajectories approach. The value of FWHM along the $x$ axis for both cases, $a_s=100a_0$ and $a_s=0$, are calculated utilizing a Voigt fit to the central focused structure for each case, and the results are illustrated respectively in Figs~\ref{f303}(e, f). For the interacting and non-interacting $^{87}$Rb BEC's, the resultant linewidth is estimated as $(\Delta x)_{\text{GPE}}^{\text{int}}=1.074~\mu$m and $(\Delta x)_{\text{GPE}}^{\text{non}}=0.477~\mu$m respectively indicating that the focal spot size for an interacting case is about two times larger than that of non-interacting BEC. Moreover, the magnitude of the central peak density in the absence of s-wave interactions is about three times greater than that of interacting condensate ($3771$ compared to $1185$  atoms/$\mu$m$^2$) showing the destructive impact of inter atomic interactions on the resolution of focused profile. At this point, one can notice that for the non-interacting case, there is a reasonable agreement between the outcomes of the classical trajectories [$(\Delta x)_{\text{Classical}}^{\text{non}}=0.495~\mu$m] and GPE [$(\Delta x)_{\text{GPE}}^{\text{non}}=0.477~\mu$m] models. However, for the interacting case, the classical model fails to accurately render an estimate for the linewidth.

\subsection{The Variation of BEC and Potential Factors}
\label{sec:11}

In this section, we consider more examples to investigate the influence of altering the BEC longitudinal velocity and lattice radius size on the deposited profiles. In addition to the previous instance discussed in section~\ref{sec:5} for the focused profile with $\sigma_z=10~\mu$m, $v_z=1$ cm/s, two more rounds of simulations are considered accounting for both interacting and non-interacting BEC's. While parameters such as $\xi=5.37$, $\lambda=12.48~\mu$m, $\Delta=200$ GHz remain unchanged, we consider $\sigma_z=20~\mu$m, $v_z=1$ cm/s and $\sigma_z=10~\mu$m, $v_z=2$ cm/s. The results for FWHM and central peak density values are summarized in Table~\ref{t301}. We notice that doubling the lattice radius size for the same BEC velocity leads to a reduction almost by half in the structure peak, and an increase by a factor of two in the structure linewidth for both $a_s=100a_0$ and $a_s=0$. Furthermore, for the two cases, increasing the condensate velocity for a given potential radius results in a decline by half in the FWHM and a growth by a factor of two in the peak density. This arises from the fact that the potential peak intensity is directly proportional to the square of BEC velocity whereas it is inversely proportional to the square of lattice radius size, $I_0\propto v_z^2/\sigma_z^2$.

\section{Velocity Distribution of a BEC}
\label{sec:6}

Due to the s-wave interactions between the atoms within a condensate, the velocity of atoms does not remain constant over time. Hence, a distribution is formed in both the longitudinal (along the $z$ axis) and transverse (along the $x$ axis) velocity profiles so that each one encompasses an associated peak representing the most probable velocity. Taking a Fast Fourier Transform (FFT) directly from the BEC density profile in position space, one is able to gain the BEC density profile in momentum space at different times of propagation. Since inter atomic interactions cause the condensate to expand (for $a_s>0$) over time the density profile in momentum space also spreads.

\begin{figure}[t!]
	\centering
	\includegraphics[width=7cm, height=12.5cm,angle=0]{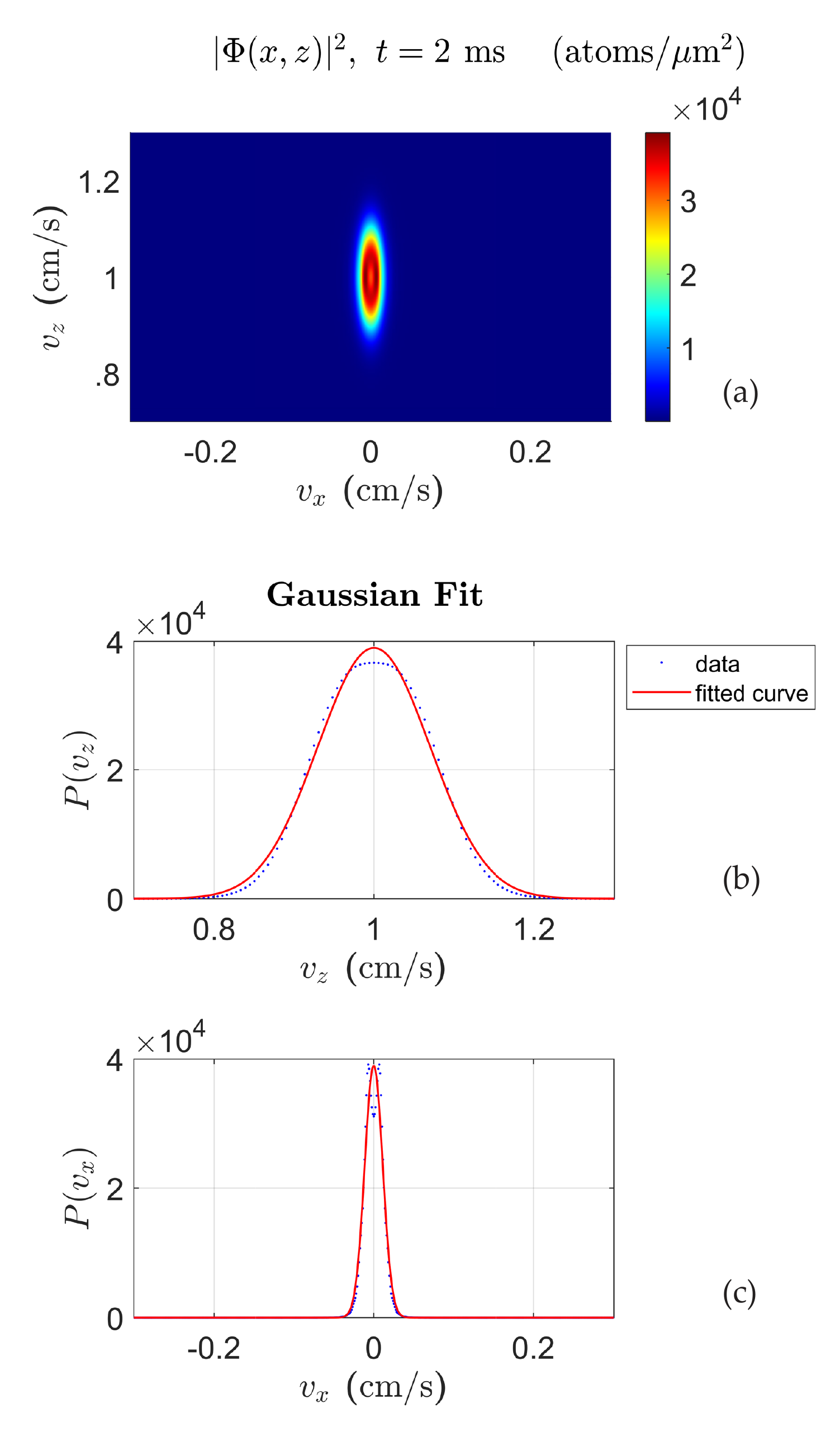}
	\caption{(a) The density profile of the free $^{87}$Rb BEC, with $a_s=100a_0$, in momentum space at $t=2$ ms. (b): The longitudinal and (c): the transverse velocity distribution profiles (shown by the blue dots) are separately taken from the BEC density profile in momentum space at $t=2$ ms. The most probable longitudinal velocity is $v_z=1$ cm/s, which is the initial velocity of the condensate (no gravity is applied to the BEC), and the peak velocity in transverse profile is $v_{x0}=0$. An appropriate Gaussian fit is applied to both plots (solid red curves), which estimates the FWHM for the longitudinal and transverse profiles as $\Delta v_z=0.1791$ cm/s and $\Delta v_x=0.0264$ cm/s respectively.}
	\label{f306}
\end{figure}

\begin{figure}[t!]
	\centering
	\includegraphics[width=9.5cm, height=4.5cm,angle=0]{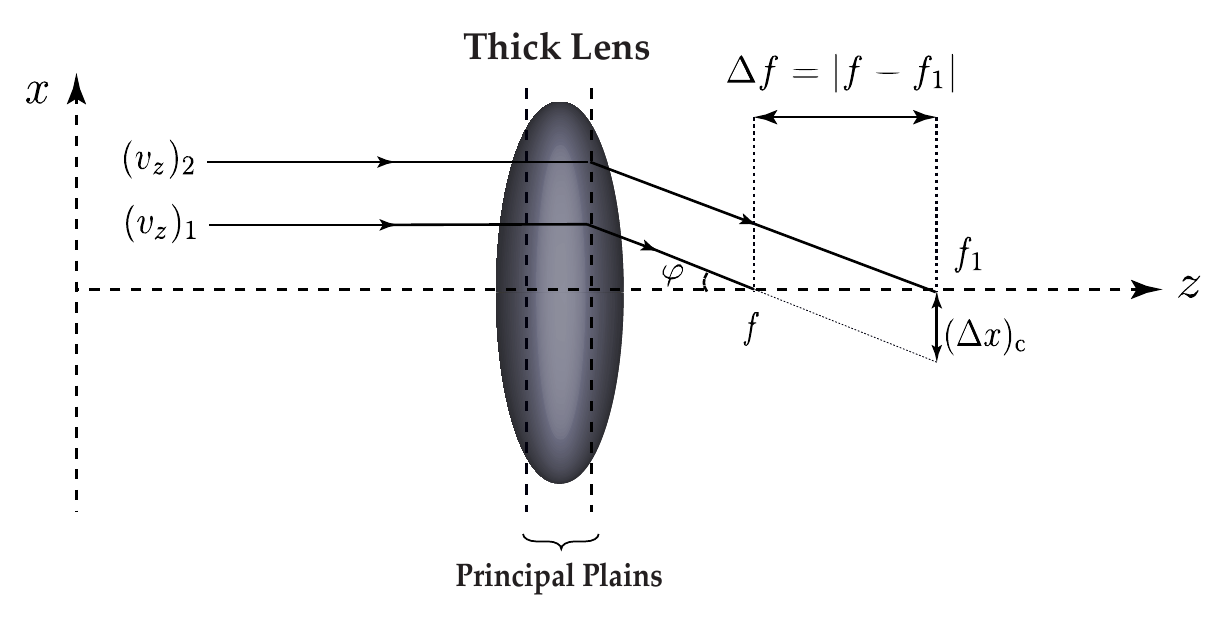}
	\caption{A schematic illustration of two atoms with different longitudinal velocities whilst crossing through a thick lens. They are then focused at different focus points due to chromatic aberration. The actual focus point, $f$, is considered for atom (1) moving at $v_{z1}$. The angle in which atom (1) creates with the focal axis ($z$ axis) is $\varphi$, $f_1$ is the focal length for atom (2)  moving at $v_{z2}$, and $(\Delta x)_{\text{c}}$ is the chromatic aberration broadening.}
	\label{f307}
\end{figure}

As an illustration, we have studied separately the BEC transverse and longitudinal velocity distributions at $t=2$ ms in Figs~\ref{f306}(a-c). Here, the BEC is kicked by $v_z=1$ cm/s when being released from the trap. Neglecting the gravity acceleration in BEC's falling, the linewidth for each distribution is derived by applying a Gaussian fit [see Figs~\ref{f306}(b, c)]. The values of FWHM for the transverse and longitudinal profiles at $t=2$ ms are estimated as $\Delta v_x=0.0264$ cm/s and $\Delta v_z=0.1791$ cm/s respectively. It is clear that the tight trap frequency along the $z$ axis ($\omega_z=2\pi\times 70$ Hz) causes a significantly enhanced widening velocity profile along the falling axis compared to the horizontal axis ($\Delta v_z\gg \Delta v_x$).

In sections~\ref{sec:7} and \ref{sec:8}, we explore the possibility of implementing the information from $\Delta v_z$ and $\Delta v_x$ to the classical trajectories model to predict the resultant structure broadenings.

\subsection{The Chromatic Aberration in Classical Trajectories Model}
\label{sec:7}

In optics, the chromatic aberration occurs because lenses have different refractive indices for different wavelengths of light causing the parallel incident wavelengths to focus at different positions from the focal point \cite{3_22}. In our case, if the longitudinal velocity of incoming atoms varies when transmitting through a focusing lens, they are not focused at a certain focal point limiting the resolution.

\begin{figure}[t!]
	\centering
	\includegraphics[width=8cm, height=7cm,angle=0]{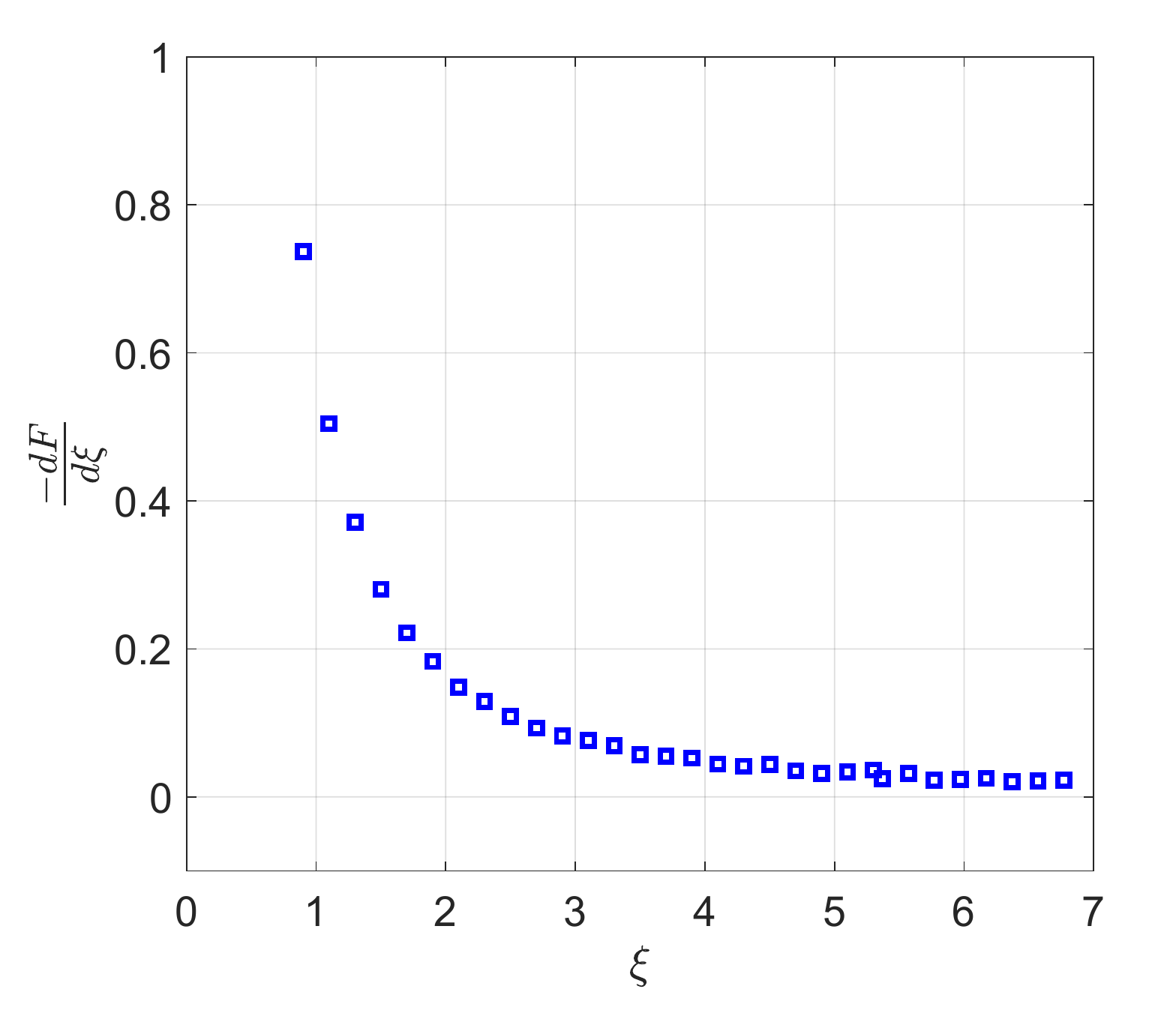}
	\caption{Variation of the dimensionless focal length, $F$, against $\xi$. This graph is used to estimate the broadening resulting from the longitudinal velocity spread in structure resolution.}
	\label{f9}
\end{figure}

In this section, the broadening contribution arising from the longitudinal velocity spread is calculated via the classical trajectories model. According to the Fig~\ref{f307}, considering the displacement of the focal length (from $f$ to $f_1$) due to the velocity variation (from $v_{z1}$ to $v_{z2}$) as $\Delta f$, and the convergence angle at the focus point as $\varphi$, the resultant broadening along the $x$ axis is given by
\begin{equation}
(\Delta x)_{\text{c}}=\varphi\Delta f,
\label{311}
\end{equation}
where 
\begin{equation}
\varphi=\tan^{-1}(D/f),
\label{312}
\end{equation}
is the angle between the incident atomic beam and the focal axis, and $D$ is the lens slit size (which is the distance between two adjacent peaks in an optical lattice) given by $D=\lambda/2$. Since a change in velocity, $v_z$, would vary the focal length, $f$, one can write 
\begin{equation}
\Delta f=\frac{df}{dv_z}\Delta v_z,
\label{313}
\end{equation}
where the variation of focal length with respect to the longitudinal velocity of atoms, $\frac{df}{dv_z}$ (or the kinetic energy of atoms) can be broken down as
\begin{equation}
\frac{df}{dv_z}=\frac{df}{d\xi}\frac{d\xi}{dv_z}.
\label{314}
\end{equation}
The dimensionless parameter, $\xi$, in Eq.~(\ref{314}) is a function of $E_0$ and consequently a function of $v_z$ [see Eq.~(\ref{5})], represented by
\begin{equation}
\xi(E_0)=\frac{\mathscr{C}}{E_0}=\frac{\mathscr{C}}{1/2mv_z^2},
\label{315}
\end{equation}
where $\mathscr{C}$ is a constant coefficient. Now, taking into account that  
\begin{equation}
\frac{d\xi}{dv_z}=-\frac{2}{v_z}\xi,
\label{316}
\end{equation}
and using Eq.~(\ref{314}), Eq.~(\ref{313}) converts to
\begin{equation}
\Delta f=-2\xi\frac{df}{d\xi}\frac{\Delta v_z}{v_z}.
\label{317}
\end{equation}
Finally, substituting Eqs.~(\ref{317}) and (\ref{312}) into Eq.~(\ref{311}), the broadening in the focal spot sizes resulting from longitudinal velocity spread is described as 
\begin{equation}
(\Delta x)_{\text{c}}=-2\xi\tan^{-1}\Big(\frac{\lambda}{2f}\Big)\frac{df}{d\xi}\frac{\Delta v_z}{v_z},
\label{318}
\end{equation}
where the data for $df/d\xi$ for a range of desired focal lengths is derived from Fig~\ref{f9} estimated by the paraxial approximation in section~\ref{sec:2}.

We now calculate the contribution of the chromatic aberration in broadening the focused structure for the example mentioned in sections~\ref{sec:2} and \ref{sec:5}. According to Fig~\ref{f9}, to focus the atoms to the center of potential, $\xi=5.37$, one would obtain $dF/d\xi=-0.0249$ ($F$ is the dimensionless focal length). Then, for a lattice with a size of $\sigma_z=10~\mu$m, one would calculate $df/d\xi=\sigma_z(dF/d\xi)=-0.249~\mu$m. Hence, given a lattice wavelength of $\lambda=12.48~\mu$m, a focal length of $f=5.531~\mu$m (for focusing to the center of the potential with $\sigma_z=10~\mu$m, and $\xi=5.37$), the longitudinal velocity and linewidth, $v_z=1$ cm/s and $\Delta v_z=0.1791$ cm/s at $t=2$ ms (for the BEC falling from $z_i=20~\mu$m to $z_f=0$), the resultant broadening magnitude is $(\Delta x)_{\text{c}}=0.405~\mu$m. 

It is clear that the most significant factor in the broadening is due to the longitudinal velocity linewidth. As a result, to reduce $(\Delta x)_{\text{c}}$, one would better choose a relatively short propagation time (or a small $z_0$) for the condensate to prevent the longitudinal velocity profile from a high spread. Moreover, from Eq.~(\ref{318}), one can understand that larger $v_z$ values can also result in less broadening in the focused structures, which is in an agreement with what is predicted by the diffraction effect [see Eq.~(\ref{226})].

\begin{figure}[t!]
	\hskip -8ex
	\includegraphics[width=9.5cm, height=4.5cm,angle=0]{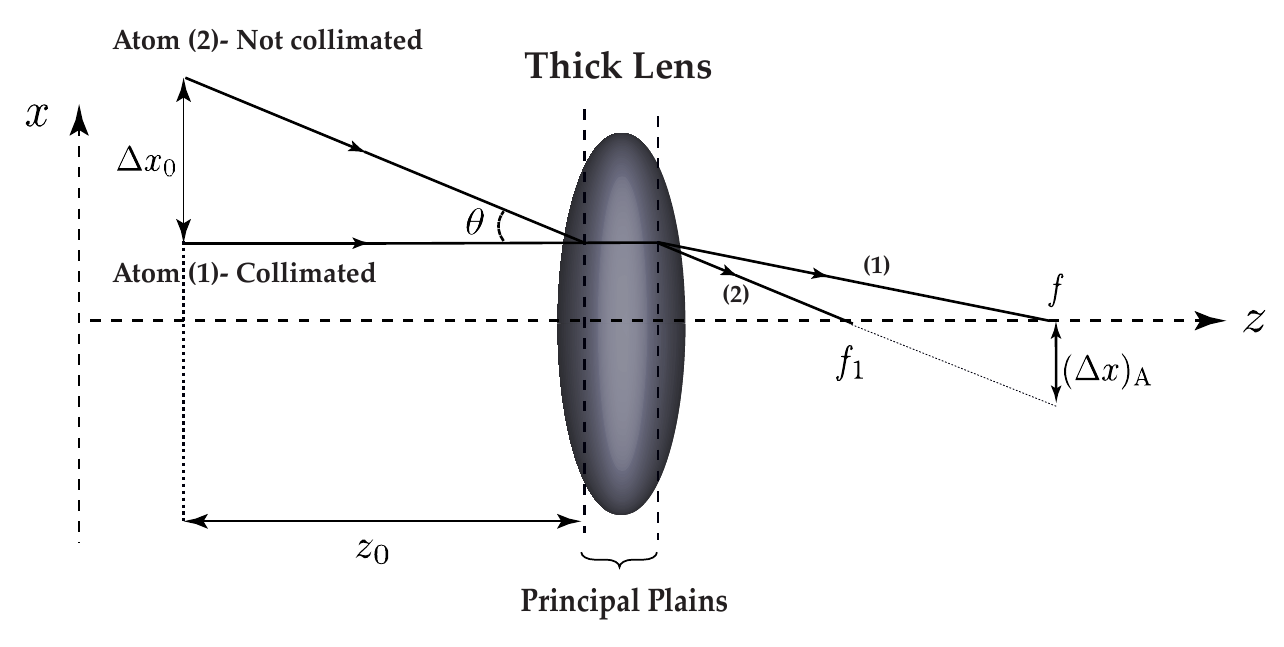}
	\caption{A schematic illustration of two atoms with an angle of $\theta$ with respect to each other crossing through a thick lens with two principal planes being focused at different points. The focal length, $f$, is considered for atom (1) collimated to the $z$ axis while $f_1$ is associated with atom (2) moving under an angle of $\theta$ with respect to the $z$ axis. The virtual object of a size of $\Delta x_0$ is located at $z=-z_0$ on the left side of the lens, and the created image of a size of $(\Delta x)_{\text{a}}$ appears on the right side of the lens.}
	\label{f308}
\end{figure}

\subsection{The Angular Divergence in Classical Trajectories Model} 
\label{sec:8}

Another broadening which appears in the deposited structures results from the a divergence in the falling beam of atoms. The divergence is sourced from the transverse velocities of the atoms within the cloud, $v_x\neq 0$. As shown in Fig~\ref{f308}, we consider two atoms approaching a thick lens. Atom (1) is collimated to the focal axis ($z$ axis), and atom (2) is moving towards the lens with an angle of $\theta$ with respect to the focal axis. The size of the virtual object caused by the two atoms located at $z=-z_0$ before the lens is chosen as $\Delta x_0$. Considering $f$ and $f_1$, respectively, as the associated focal lengths to atom (1) and (2), the created image is demagnified by $(\Delta x)_{\text{a}}$. This is represented by the following equation

\begin{equation}
\frac{(\Delta x)_{\text{a}}}{\Delta x_0}=\frac{f}{z_0}.
\label{319}
\end{equation}
Given that the object size is obtained by
\begin{equation}
\Delta x_0=z_0\theta, 
\label{320}
\end{equation}
the linewidth arising from the beam divergence would be estimated as
\begin{equation}
(\Delta x)_{\text{a}}=f\theta,
\label{321}
\end{equation}
where the beam divergence, 
\begin{equation}
\theta=\tan^{-1}(\Delta v_x/v_z),
\label{322}
\end{equation}
is directly taken from the transverse velocity linewidth, $\Delta v_x$, and the most probable longitudinal velocity, $v_z$. Inserting Eq.~(\ref{322}) into Eq.~(\ref{321}), one would obtain
\begin{equation}
(\Delta x)_{\text{a}}=f\tan^{-1}(\Delta v_x/v_z).
\label{323}
\end{equation}

As a case in point, we refer to the example in sections~\ref{sec:2} and \ref{sec:5}. A transverse velocity linewidth of $\Delta v_x=0.0264$ cm/s [see Fig.~\ref{f306}(c)] and a probable longitudinal velocity of $v_z=1$ cm/s result in $\theta=0.0264$ rad at $t=2$ ms for the BEC falling from $z_i=20~\mu$m focusing to the center of lattice ($\xi=5.37$) of a radius size of $\sigma_z=10~\mu$m. In such an instance, considering the associated focal length, $f=5.531~\mu$m, the angular divergence broadening is estimated as $(\Delta x)_{\text{a}}= 0.146~\mu$m. 

Clearly, a longer time of BEC propagation gives the atoms more chance to interact within the cloud. Thus, to reduce $\Delta v_x$ and consequently $(\Delta x)_{\text{a}}$, one can choose to utilize a fall time for the focusing process.

\section{GPE and Classical Trajectories Agreement} 
\label{sec:9}

As stated thus far, the influence of the s-wave interaction between atoms appears as the longitudinal and transverse velocity distributions in the classical calculations. That being said, one can define the following equivalency

\begin{equation}
(\Delta x)^{\text{int}}_{\text{GPE}}\equiv(\Delta x)^{\text{int}}_{\text{Classical}},
\label{324}
\end{equation}
where 
\begin{equation}
(\Delta x)^{\text{int}}_{\text{Classical}}=(\Delta x)^{\text{non}}_{\text{Classical}}+(\Delta x)_{\text{c}}+(\Delta x)_{\text{a}},
\label{325}
\end{equation}
which can be rewritten as following using Eq.~(\ref{227}),
\begin{equation}
(\Delta x)^{\text{int}}_{\text{Classical}}=(\Delta x)_{\text{sph}}+(\Delta x)_{\text{diff}}+(\Delta x)_{\text{c}}+(\Delta x)_{\text{a}}.
\label{326}
\end{equation}

For the examples discussed in sections~\ref{sec:7} and \ref{sec:8}, adding $(\Delta x)_\text{c}=0.405~\mu$m and $(\Delta x)_{\text{a}}=0.146~\mu$m to $(\Delta x)^{\text{non}}_{\text{Classical}}=0.495~\mu$m [which comprises $(\Delta x)_{\text{sph}}=0.136~\mu$m and $(\Delta x)_{\text{diff}}=0.359~\mu$m, see section~(\ref{sec:2})], we estimate the total FWHM of the focused structures as $(\Delta x)^{\text{int}}_{\text{Classical}}=1.046~\mu$m via the classical trajectories model. This is in good agreement with GPE results, $(\Delta x)^{\text{int}}_{\text{GPE}}=1.074~\mu$m [see Fig.~\ref{f303}(e)] indicating a reliable mapping between the classical trajectories approach using the BEC transverse and longitudinal velocity profiles from the GPE. Table~\ref{t302} compares the outcomes resulting from the GPE and classical methods for three various examples for both interacting and non-interacting cases.

\begin{table}
	\centering
	\footnotesize
	\begin{tabular}{|c|cc|cc|} 
		\hline
		Example & \parbox[t]{1.5cm}{\raggedright $(\Delta x)_{\text{Class}}^{\text{non}}$} & \parbox[t]{1.5cm}{\raggedright $(\Delta x)_{\text{GPE}}^{\text{non}}$} & \parbox[t]{1.5cm}{\raggedright $(\Delta x)_{\text{Class}}^{\text{int}}$} & \parbox[t]{1.5cm}{\raggedright $(\Delta x)_{\text{GPE}}^{\text{int}}$}\\
		
		\hline
		\parbox[t]{2cm}{\raggedright $\sigma_z=10~\mu$m $v_z=1$ cm/s} & $0.495$ ~~~~& $0.477$ & $1.046$~~~~ & $1.074$ \\
		\hline
		\parbox[t]{2cm}{\raggedright $\sigma_z=20~\mu$m $v_z=1$ cm/s}  & $1.059$~~~~ & $0.985$ & $2.065$ ~~~~& $2.106$ \\
		\hline
		\parbox[t]{2cm}{\raggedright $\sigma_z=10~\mu$m $v_z=2$ cm/s} & $0.264$ ~~~~& $0.228$ & $0.428$ ~~~~& $0.447$ \\
		\hline
	\end{tabular}
	\caption{FWHM data in $\mu$m calculated via the GPE and classical trajectories approaches for interacting ($a_s=100a_0$) and non-interacting ($a_s=0$) focused BEC's collected at $z=0$ for three different simulations \{$\sigma_z=10~\mu$m, $v_z=1$ cm/s\}; \{$\sigma_z=20~\mu$m, $v_z=1$ cm/s\} and \{$\sigma_z=10~\mu$m, $v_z=2$ cm/s\}.}
	\label{t302}
\end{table}

\section{Numerical investigation of focusing}
\label{sec:10}

We now study in more detail the impact of the focusing potential aperture size (or the potential wavelength) as well as its radius on the deposited BEC structures. We consider a $^{87}$Rb BEC trapped by $\omega_x=2\pi\times 10$ Hz, and $\omega_y=\omega_z=2\pi\times 70$ Hz, whose center-of-mass is at $z_0=500~\mu$m from the lattice center at $z=0$. In the results presented, the freely propagating and interacting ($a_s=100a_0$) condensate is aimed to be focused optimally at $z=0$ ($\xi=5.37$). 

\begin{figure}[t!]
	\hskip -8ex
	\includegraphics[width=9.5cm, height=12.5cm,angle=0]{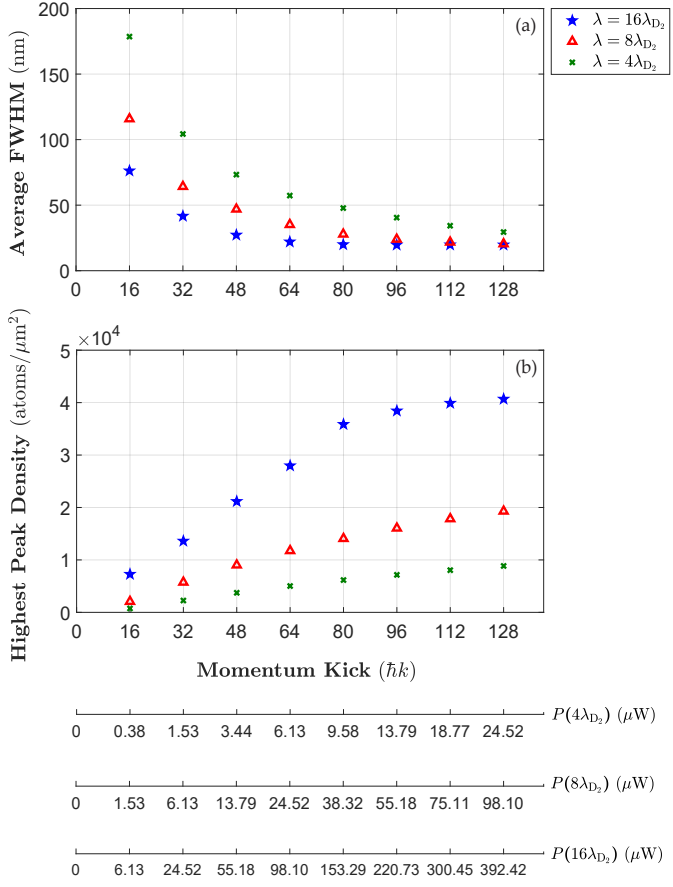}
	\caption{Results for the characteristic factors of the focused $^{87}$Rb BEC versus different momentum kicks. (a, b): Results for the average structure resolutions and the highest peak densities. The data points indicated by the blue stars, red triangles and green crosses are for $\lambda=16$, $8$ and $4\lambda_{\text{D}_2}$ respectively. The three horizontal axes at the bottom of the figure represent the corresponding optimal power values to the momentum kicks for each lattice wavelength. Parameters involved in the simulations are: $N=10^5$, $z_0 =500~\mu$m, $\sigma_z=10~\mu$m, $a_s=100a_0$ , $\Delta=200$ GHz, $\gamma=37$ MHz, and $I_s=16.5$ W/m$^2$.}
	\label{f309}
\end{figure}

We initially consider a lattice radius size of $\sigma_z=10~\mu$m while three various potential wavelengths are employed, $\lambda=16\lambda_{\text{D}_2}$, $8\lambda_{\text{D}_2}$, and $4\lambda_{\text{D}_2}$. The results are extracted for a range of initial momentum kicks differing from $p=16~\hbar k$ to $p=128~\hbar k$ (where $k$ is the wavenumber associated with the $^{87}$Rb D$_2$ line defined by $k=2\pi/\lambda_{\text{D}_2}$) is applied to the BEC. As a result, for an optimal focus, the lattice power and peak intensity are adjusted accordingly based on the BEC kinetic energy for any certain momentum kick. Figs.~\ref{f309}(a, b) display the associated outputs for the focused BEC linewidths and peak densities respectively. The FWHM in each numerical calculation is achieved by taking an average over the linewidth of the structure peaks whose height are greater than $1/e$ of the highest central peak.

\begin{figure}[t!]
	\hskip 8ex
	\includegraphics[width=9.5cm, height=12.5cm,angle=0]{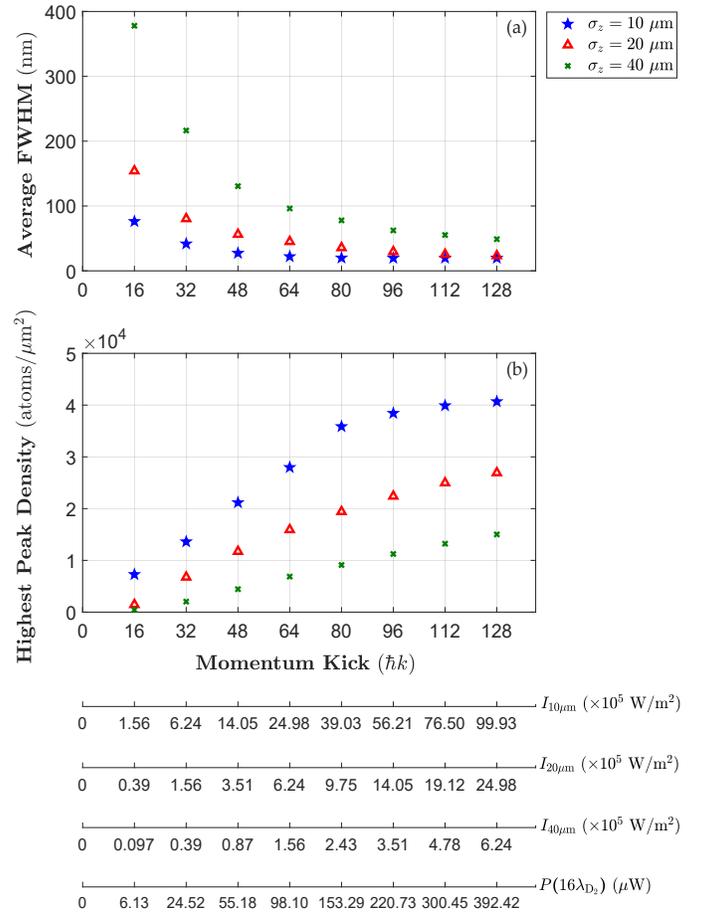}
	\caption{Results for the characteristic factors of the focused $^{87}$Rb BEC versus different momentum kicks. (a, b): Results for the average structure resolutions and the highest peak densities. The data points indicated by the blue stars, red triangles and green crosses are for $\sigma_z=10$, $20$ and $40~\mu$m respectively. The first three horizontal axes at the bottom of the figure represent the corresponding peak intensity values to the momentum kicks for each lattice radius size. The fourth horizontal axis displays the corresponding optimal power values for $\lambda=16\lambda_{\text{D}_2}$. Parameters involved in the simulations are: $N=10^5$, $z_0 =500~\mu$m, $\lambda=12.48~\mu$m, $a_s=100a_0$ , $\Delta=200$ GHz, $\gamma=37$ MHz, and $I_s=16.5$ W/m$^2$.}
	\label{f310}
\end{figure}

Analyzing the results, firstly, one can conclude that imparting higher momentum kicks to the BEC reduces the focal spot sizes and enhances the peak densities for any potential wavelength. However, larger $\lambda$ values necessitate higher potential powers and peak intensities resulting in the superior structure resolutions and heights (see the blue stars in Figs.~\ref{f309}(a, b) compared to the red triangles and green crosses). Furthermore, one can understand that the FWHM and peak density data points for $\lambda=16\lambda_{\text{D}_2}$, $8\lambda_{\text{D}_2}$, and $4\lambda_{\text{D}_2}$ tend to a steady state at relatively large magnitudes of initial momentum kick [i.e. $p\geq 96~\hbar k$, see Figs.~\ref{f309}(a, b)]. This implies the  characteristic factors of a focused profile (including the resolution and peak density) become independent of the BEC momentum kick at relatively high longitudinal velocities. At this point, the profile resolution also becomes independent of the lattice wavelength while the focused peak density remains strongly dependent on the size of the momentum kick. 

\begin{figure}[t!]
	\hskip -8ex
	\includegraphics[width=9.5cm, height=11cm,angle=0]{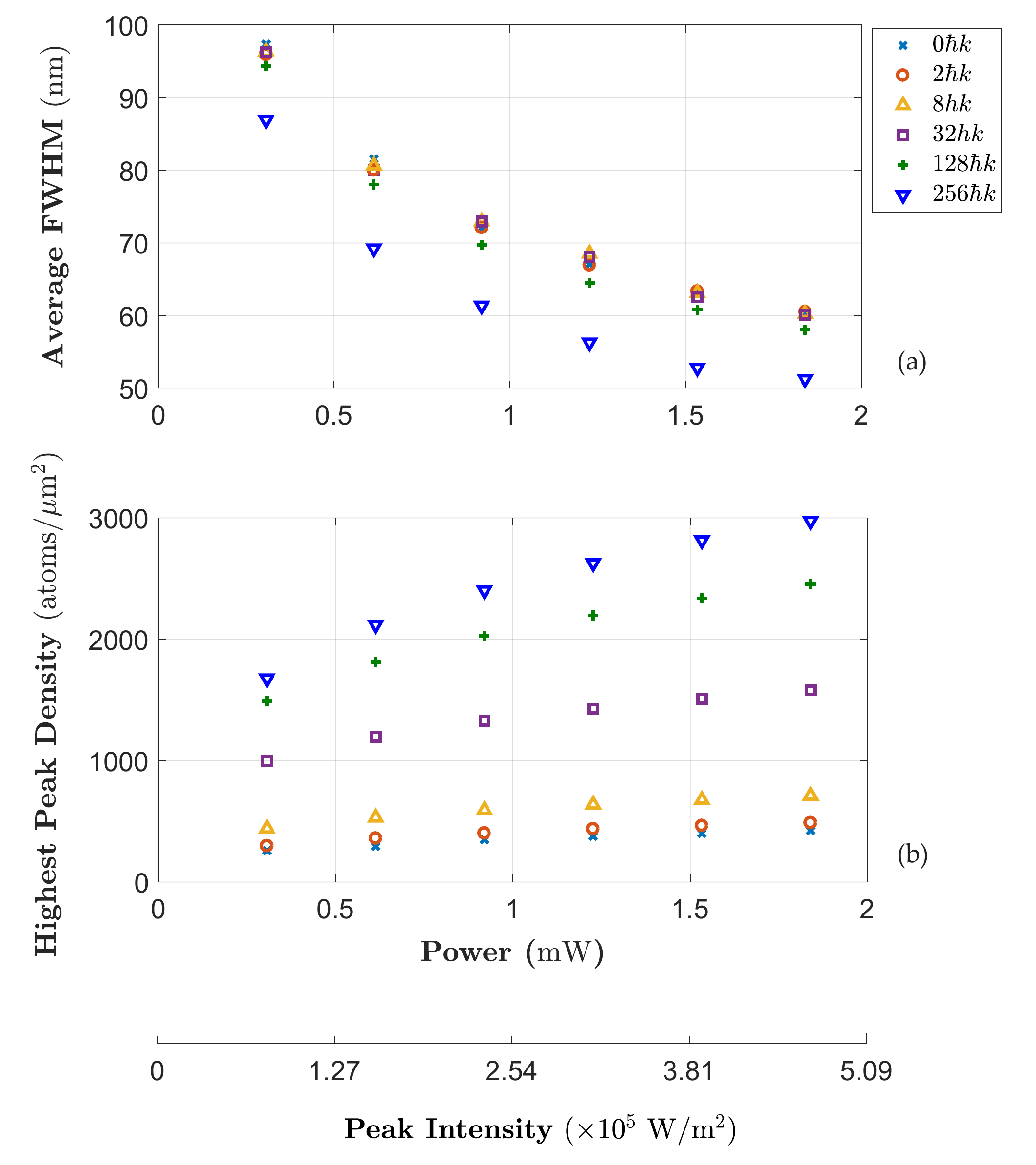}
	\caption{Results for the characteristic factors of the focused $^{87}$Rb BEC versus different potential power values for $\lambda=\lambda_{\text{D}_2}$. (a, b): Results for the average structure resolutions and the highest peak densities. The data points correspond to the momenta kicks indicated in the legend. The horizontal axis at the bottom of the figure represents the corresponding peak intensity values to the momentum kicks for $\sigma_z=100~\mu$m. Parameters involved in the simulations are: $N=10^5$, $z_0 =500~\mu$m, $\lambda=780.027~\mu$m, $a_s=100a_0$ , $\Delta=200$ GHz, $\gamma=37$ MHz, and $I_s=16.5$ W/m$^2$.}
	\label{f311}
\end{figure}

\begin{figure}[t!]
	\hskip 8ex
	\includegraphics[width=9.5cm, height=11cm,angle=0]{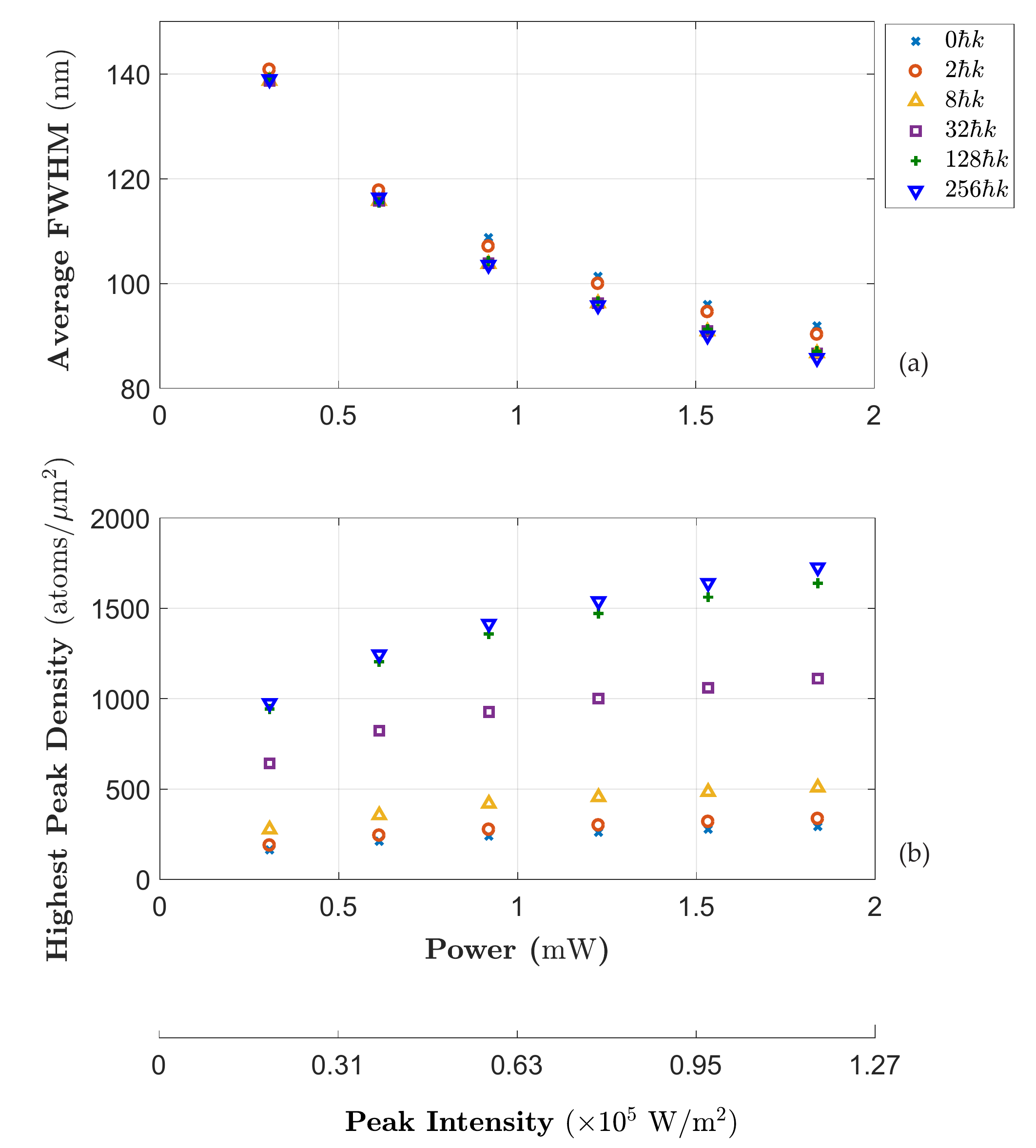}
	\caption{Results for the characteristic factors of the focused $^{87}$Rb BEC versus different potential power values for $\lambda=\lambda_{\text{D}_2}$. (a, b): Results for the average structure resolutions and the highest peak densities. The data points correspond to the momenta kicks indicated in the legend. The horizontal axis at the bottom of the figure represents the corresponding peak intensity values to the momentum kicks for $\sigma_z=200~\mu$m. Parameters involved in the simulations are: $N=10^5$, $z_0 =500~\mu$m, $\lambda=780.027~\mu$m, $a_s=100a_0$ , $\Delta=200$ GHz, $\gamma=37$ MHz, and $I_s=16.5$ W/m$^2$.}
	\label{f312}
\end{figure}

In Fig.~\ref{f310}(a, b), a fixed value of $\lambda=16\lambda_{\text{D}_2}$ is considered while three different lattice radii, $\sigma_z=10$, $20$ and $40~\mu$m are selected. Since an expansion in the potential radius size leads to a decline in the power and peak intensity values, it is expected larger FWHMs will be obtained as well as lower peak densities for $\sigma_z=40~\mu$m (indicated by the green crosses) compared to those of $\sigma_z=10~\mu$m (indicated by the blue stars). As an instance, choosing $\sigma_z=10$, $20$ and $40~\mu$m would respectively result in $(\Delta x)_{\text{GPE}}^{\text{int}}=19.68$, $23.27$ and $48.83$ nm for the resolutions, and $4.068\times 10^4$, $2.692\times 10^4$ and $1.501\times 10^4$ atoms/$\mu\text{m}^2$ for the peak densities at $p=128~\hbar k$. Moreover, the profile characteristic factors tend to a steady state for every potential size at larger momentum kicks [i.e. $p\geq 96~\hbar k$, see Figs.~\ref{f310}(a, b)].

Finally, the focused $^{87}$Rb BEC profile is studied through a focusing lattice of $\lambda= \lambda_{\text{D}_2}$. The output FWHM and peak density values are plotted as a function of potential power in Figs.~\ref{f311}(a, b) [for $\sigma_z=100~\mu$m] and Figs.~\ref{f312}(a, b) [for $\sigma_z=200~\mu$m]. There are six categories of data points representing the condensate kicked by, $p=0$, $2$, $8$, $32$, $128$ and $256~\hbar k$. Unlike the previous case, the lattice power selection here takes a certain range (from $P=0.3$ to $1.83$ mW) regardless of an optimal power value for any momentum kick, $P=P_{\lambda_{\text{D}_2}}(n\hbar k)$. In other words, the focused profiles are examined at $z=0$ even when they are not at their optimal focus stage.

According to Figs.~\ref{f311}(a, b), for each value of momentum kick, a rise in the focusing potential power results in an enhanced resolution as well as a higher peak density. For instance, setting $P=1.83$ mW for $p=256~\hbar k$ produces a focused profile with $(\Delta x)_{\text{GPE}}^{\text{int}}=51.25$ nm and a highest peak of $2975$ atoms/$\mu\text{m}^2$. However, increasing the value of beam radius (from $\sigma_z=100$ to $200~\mu$m) broadens the corresponding profile resolutions creating shorter peaks, which is indicated in Figs.~\ref{f312}(a, b). In such a case, for $P=1.83$ mW and $p=256~\hbar k$, the resolution and highest peak density respectively reduce to $(\Delta x)_{\text{GPE}}^{\text{int}}=85.87$ nm and $1726$ atoms/$\mu\text{m}^2$.

It is also noticeable that for any magnitude of momentum kick applied to the BEC, the focused profile characteristic factors become independent of the potential power at larger $P$ values. As a result, in this case, there exists a threshold in power value for improving the profile resolution and peak density meaning that one may not expect to obtain a considerable improvement in focal spot sizes when using $P>2$ mW [see Figs.~\ref{f311}(a, b) and Figs.~\ref{f312}(a, b)]. We note that the threshold range, $P>2$ mW, is only the case for a lattice of $\lambda= \lambda_{\text{D}_2}$ as it could be reduced by a factor of ten to $P_{16\lambda_{\text{D}_2}}(128~\hbar k)=0.39$ mW for a lattice of $\lambda=16\lambda_{\text{D}_2}$, which could lead to a focused profile of $(\Delta x)_{\text{GPE}}^{\text{int}}=19.68$ nm [see Fig.~\ref{f310}(a)].

\section{Conclusions}

In this paper, we summarized the classical trajectories approach and its applications in focusing ultra-cold $^{87}$Rb atoms. The structure of an appropriate focusing potential as well as its focal properties for an optimal focus were analyzed using the paraxial solution to the equations of atomic motion. Moreover, we derived a relation between the required value of potential power and the desired focus point on the focal plane along with employing a relevant example. The spherical aberration on broadening of focused structures was explored through numerical solutions to the classical equations. We then calculated the structure linewidth resulting from the diffraction contribution. We inferred that an increase in lattice aperture size and atomic longitudinal velocity or a reduction in the focal length results in a lower FWHM of diffraction as in this case atoms would have less chance to interfere together. \\

We then moved on to consider the influence of s-wave interactions between the atoms on focusing of $^{87}$Rb condensate using the GPE. It is found that the resultant structure linewidths are significantly impacted by s-wave interactions. Moreover, we concluded that either reducing the lattice radius size or increasing the BEC initial kinetic energy could improve the characteristic factors (resolutions and peak densities) of a focused $^{87}$Rb profile. Further to this, the contribution for both the BEC longitudinal and transverse velocity distributions in broadening the structure size was explored via the classical trajectories model. We found a reliable agreement between the GPE approach and a classical trajectories model which includes the interaction contribution. \\

Finally, conducting a variety of numerical simulations with different lattice radii and wavelengths, we noticed that narrower and higher density structures arise from relatively lower potential radius sizes and greater wavelengths. It was shown that nano-meter structures are achievable in principle. The example of this is a resultant structure of $(\Delta x)_{\text{GPE}}^{\text{int}}=19.68$ nm via a focusing lattice of $\sigma_z=10~\mu$m and $\lambda=16\lambda_{\text{D}_2}$. Furthermore, it was inferred that applying higher momentum kicks necessitating greater optimal potential powers and peak intensities cause superior profile resolutions and peak fluxes. However, for a certain lattice wavelength, there is a threshold point in potential power where for larger values than this point, the focused profile characteristic factors are no longer dependent on power magnitudes tending to a steady state. It was observed that exploiting a relatively smaller lattice wavelength increases the threshold power whilst causing a destructive impact on the structure resolution and peak density. For instance, the approximate threshold power for a lattice of $\lambda=\lambda_{\text{D}_2}$ and $\lambda=16\lambda_{\text{D}_2}$ is about $P_{\lambda_{\text{D}_2}}=1.83$ mW and $P_{16\lambda_{\text{D}_2}}=0.39$ mW resulting in $(\Delta x)_{\lambda_{\text{D}_2}}^{\text{GPE}}=51.25$ nm and $(\Delta x)_{16\lambda_{\text{D}_2}}^{\text{GPE}}=19.68$ nm.

\section*{ACKNOWLEDGMENTS}

The authors would like to thank Nicholas P. Robins for useful discussions and feedback. Grateful acknowledgement is also extended to Timothy Senden for the financial support of the project. This research was supported by the Research School of Physics, the Australian National University, and the School of Physics, University of Melbourne.

\bibliography{Ref_1,Ref_2,Ref_3,Ref_4,Ref_5,Ref_6,Ref_7,Ref_8,Ref_9,Ref_10,Ref_11,Ref_12,Ref_13,Ref_14,Ref_15,Ref_16,Ref_17,Ref_18,Ref_19,Ref_20,Ref_21,Ref_22,Ref_23,Ref_24,Ref_25,Ref_26,Ref_27,Ref_28,Ref_29,Ref_30,Ref_31,Ref_32,Ref_33,Ref_0,RRef_1,RRef_2,RRef_3,RRef_54,RRef_44,RRef_45,RRef_46,RRef_49,RRef_50,RRef_51,RRef_52,RRef_7,RRef_8,RRef_9,RRef_4,RRef_5,RRef_20,RRef_10,RRef_11,RRef_12,RRef_13,RRef_53,R2_13,R2_14,R2_15,R2_16,R3_1,R3_2,R3_3,R3_4,R3_5,R3_6,R3_7,R3_8,R3_9,R3_10,R3_11,R3_12,R3_13,R3_14,R3_15,R3_16,R3_17,R3_18,R3_19,R3_20,R3_21,R3_22}
\end{document}